\documentclass[a4paper,11pt]{article}
\pdfoutput=1
\usepackage{jcappub}
\usepackage[T1]{fontenc}
\usepackage[utf8]{inputenc}
\usepackage{aas_macros}
\usepackage[normalem]{ulem}
\DeclareUnicodeCharacter{2212}{-}

\usepackage{amsmath,amssymb,cleveref,natbib}
\usepackage[version=4]{mhchem}
\usepackage{siunitx}
\DeclareSIUnit\year{yr}

\newcommand\mpl{
    M_{\mathrm{Pl}}
}
\newcommand\mstop{
    M_{\mathrm{CPR}}
}
\newcommand\mfreeze{
    M_{\mathrm{min}}^{(e)}
}
\newcommand\fcpr{
    f_{\mathrm{CPR}}
}
\newcommand\order{
    \mathcal O
}

\newcommand\du{
    \mathrm{d}
}
\newcommand\dd{
    \,\du
}
\newcommand\SRIM{\texttt{SRIM}}
\newcommand\detectortype[1]{
    \par\medskip\noindent\textit{#1. }
}

\title{Direct detection of primordial \\ black hole relics as dark matter}
\author{Benjamin V. Lehmann,}
\author{Christian Johnson,}
\author{Stefano Profumo,}
\author{and Thomas Schwemberger}
\affiliation{
    Department of Physics, University of California, Santa Cruz\\
    1156 High Street, Santa Cruz, CA 95064, USA
}
\affiliation{
    Santa Cruz Institute for Particle Physics\\
    1156 High Street, Santa Cruz, CA 95064, USA
}
\emailAdd{blehmann@ucsc.edu}
\emailAdd{arcjohns@ucsc.edu}
\emailAdd{profumo@ucsc.edu}
\emailAdd{tschwemb@ucsc.edu}

\abstract{
    If dark matter is composed of primordial black holes, such black holes can span an enormous range of masses. A variety of observational constraints exist on massive black holes, and black holes with masses below \SI{e15}{\gram} are often assumed to have completely evaporated by the present day. But if the evaporation process halts at the Planck scale, it would leave behind a stable relic, and such objects could constitute the entirety of dark matter. Neutral Planck-scale relics are effectively invisible to both astrophysical and direct detection searches. However, we argue that such relics may typically carry electric charge, making them visible to terrestrial detectors. We  evaluate constraints and detection prospects in detail, and show that if not already ruled out by monopole searches, this scenario can be largely explored within the next decade using existing or planned experimental equipment. A single detection would have enormous implications for cosmology, black hole physics, and quantum gravity.
}
\keywords{
    primordial black holes, dark matter theory, dark matter experiments, dark matter detectors
}

\begin{document}
\maketitle
\flushbottom

\section{Introduction}
The last few years have seen a resurgence of interest in black holes as a dark matter candidate. Many of the simplest and best-motivated \emph{particle} dark matter candidates have been strongly constrained by a combination of collider searches \citep{Kahlhoefer:2017dnp} and direct detection experiments \citep{Agnese:2017njq,Aprile:2018dbl,Cui:2017nnn,Arcadi:2017kky}. Macroscopic objects such as black holes are an attractively minimal dark matter candidate by comparison. Many scenarios predict the production of primordial black holes (PBHs) in the early universe without the addition of any new fields. Such black holes could constitute a significant fraction of cosmological dark matter \citep{Carr:2016drx}, and could simultaneously serve as a probe of high-scale physics which sets the conditions of their formation \citep{Carr:2003bj}.

The detection of gravitational waves by LIGO spurred additional enthusiasm for the possibility that PBHs could account for dark matter, with many analyses of PBH dark matter at masses of order $50M_\odot$ \citep{Abbott:2016blz}. However, interest in PBH dark matter has been reignited across the mass spectrum. Substantial efforts have been made over the last few years to detect PBHs at all scales, and many constraints limit the fraction of dark matter in PBHs with masses above $\sim10^{-6}M_\odot$. For smaller black holes, there remain mass windows in which PBH-dominated dark matter is viable. In particular, microlensing constraints have recently been relaxed at masses for which the Schwarzschild radius is of order the wavelength of light surveyed \citep{Niikura:2017zjd, Sugiyama:2019dgt}. But even in this regime, constraints will strengthen with additional data and modeling of these finite-size effects.

This leaves the extremely light black holes, i.e., those produced with a mass below $\sim\SI{e14}{\gram}$. Such small black holes are expected to be unstable due to Hawking radiation: they should completely evaporate within the lifetime of the universe. The evaporation process has been used to draw constraints on the population of light black holes today \citep{Carr:2009jm,Archambault:2017asc,Johnson:2019koa}.  However, evaporation is not well-understood at masses of order the Planck scale. It has been suggested that Hawking radiation in fact halts near this scale, leaving a relic black hole of mass $\sim\mpl$ \citep{Chen:2002tu,Alexeyev:2002tg,Chen:2004ft,Nozari:2005ah}, and these relics could constitute the entirety of dark matter \citep{MacGibbon:1987my,Barrow:1992hq,Carr:1994ar}. Such a relic would be almost completely inert, interacting only via gravity, but with a mass far too small to be detected as an individual object. From an experimental viewpoint, dark matter in the form of Planck-scale relics is a ``nightmare'' scenario, in that dark matter is effectively a particle with no non-gravitational interaction with the standard model. As such, it is extremely difficult to constrain relic black holes as dark matter.

However, there is another possibility: suppose that such relic black holes were electrically charged. Then these objects might be detectable by existing means. Interestingly, as we will discuss here, there is reason to believe that relic black holes could \emph{typically} carry non-zero charge. The scenario is as follows: as the black hole evaporates, it emits charged particles of both signs, and it does so stochastically. Thus, during the evaporation process, non-zero electric charges are generic. If evaporation is cut off sharply at some mass scale of order $\mpl$, the black hole might be frozen with leftover electric charge of random sign. Alternatively, as we will also  discuss, the impact of the spontaneous charge itself on the black hole geometry may act as a stabilizing mechanism. Regardless of their origin, we call such objects \emph{Charged Planck-scale Relics} (CPRs). In this work, we show that such objects, if they exist,  would be detectable terrestrially.

Generally, electric charges of order $e$ are considered to be incompatible with dark matter. However, experimental constraints on the charge of dark matter (e.g. in the context of millicharged dark matter) are always placed on some combination of the charge and mass of the dark matter species. In our case, we will be interested in objects with a charge-to-mass ratio of order $\sim e/\mpl$. Such objects behave as dark matter in every respect: their self-interactions are dominated by gravity; their interactions with standard model particles impart no appreciable change in their momentum; and, since they must be extremely sparse due to their large masses, they have no impact on baryonic dynamics apart from their bulk gravitational potential.

CPRs are similar to charged massive particles (CHAMPs \citep{DeRujula:1989fe}) in that they possess integer-valued electric charges. CHAMPs have been studied as a dark matter candidate for decades, but direct detection prospects differ significantly between CHAMPs and CPRs, due mainly to the difference in the typical masses of the two objects. CHAMPs are depleted in the galactic disk due to their interactions with magnetic fields \citep{Chuzhoy:2008zy,McDermott:2010pa}, and a survey of other CHAMP probes by \citet{Dimopoulos:1989hk} yielded null results. However, these results apply only to CHAMPs with masses \emph{below} \SI{e8}{\tera\electronvolt}. We expect CPRs to be found at $\sim\SI{e16}{\tera\electronvolt}$, well above this threshold, so the CHAMP literature is largely inapplicable to our case.

The strongest cosmological constraint on the charge-to-mass ratio of dark matter comes from the CMB power spectrum \citep{Dolgov:2013una}, which requires
\begin{equation}
    q_{\mathrm{DM}} \lesssim 2.24\times10^{-4}\left(\frac{m_{\mathrm{DM}}}{\SI{1}{\tera\electronvolt}}\right)^{1/2}e.
\end{equation}
Our fiducial mass scale is $\mpl$, for which this translates to $q_{\mathrm{DM}}\lesssim2.5\times10^3e$. This constraint is thus also irrelevant for our scenario, in which, as detailed below, we predict charges of order $e$. Indeed, as we will discuss, the cosmic censorship conjecture imposes a much stronger constraint on the electric charge of Planck-scale black holes. Constraints on $q_{\mathrm{DM}}$ from terrestrial experiments are also ineffective at the large masses we consider.

Thus, there are two major motivations to search for CPRs experimentally. First, despite being electrically charged, CPRs could constitute the entirety of dark matter if evaporation halts near the Planck scale. Second, even if CPRs constitute only a small fraction of dark matter, the confirmed detection of even one such object would be of incredible value to black hole physics: it would confirm that black hole evaporation does indeed halt, and pave the way for the experimental study of quantum gravity. Remarkably, the first constraints on the abundance of CPRs can already be placed with existing experimental results, and future experiments offer the opportunity to considerably tighten these bounds.

The structure of this work is as follows. In \cref{sec:evaporation}, we show how CPRs can form, and quantitatively estimate their abundance given realistic formation scenarios. In \cref{sec:detection}, we study the interaction of CPRs with matter and evaluate mechanisms for the terrestrial detection of CPRs. In \cref{sec:prospects}, we derive constraints from non-detection in existing experiments and project constraints that can be obtained from proposed or upcoming experiments. We discuss the implications in \cref{sec:discussion} and conclude in \cref{sec:conclusions}.

Unless otherwise indicated, we work in units with $c=\hbar=k_B=G=1$, and $\varepsilon_0=1/4\pi$. In these units, the elementary charge $e$ is given by $\sqrt\alpha\approx 1/11.7$. We take $\mpl=\left(\hbar c/G\right)^{1/2}=1$. In these units, a black hole with charge-to-mass ratio $Q/M$ has $Q\approx (Q/M)(11.7e)$. Additionally, note that $e$ corresponds to ``positive charge'' in these units.

\section{Evaporation and spontaneous charge}
\label{sec:evaporation}
\citet{Hawking:1974} showed that black holes radiate, or \emph{evaporate}, 
as thermal blackbodies. A black hole's temperature is related to its surface gravity $\kappa$ via $T=\kappa/(2\pi)$, and according to no-hair theorems \citep{Israel:1967wq}, $\kappa$ can only depend on three parameters: the black hole's mass $M$, electric charge $Q$, and angular momentum $L$. As a benchmark, a Schwarzschild black hole ($Q=L=0$) of mass $M$ has temperature $T = 1/(8\pi M)$ as measured by a faraway observer. Since evaporation tends to discharge angular momentum rapidly, a black hole with some initial spin is unlikely to have appreciable angular momentum once it reaches the Planck scale. In particular, \citet{Page:1976b} showed that black holes with mass below $\sim\SI{e14}{\gram}$ today should have a spin parameter very near zero, so the impact of angular momentum on the black hole metric should be negligible.

Thus, we will only consider non-rotating ($L=0$) black holes with charge $Q$. Such black holes are described by the Reissner-Nordstr\"om (RN) metric:
\begin{equation}
    \du s^2 = \left(1 + \frac{2M}{r} + \frac{Q^2}{r^2}\right) \dd t^2
        - \left(1 - \frac{2M}{r} + \frac{Q^2}{r^2}\right)^{-1} \dd r^2
        - r^2 \dd^2\Omega.
\end{equation}
The radial component of the RN metric diverges at two values of $r$, namely
\begin{equation}
    r_\pm = M \pm \sqrt{M^2-Q^2}.
\end{equation}
The outer horizon radius $r_+$ defines the surface of the black hole for our purposes, and thus plays an important role in determining the properties of particle emission. Note that we only have two distinct horizons when $Q<M$. When $Q=M$, the black hole is extremal, and its surface gravity vanishes. If $Q>M$, the black hole is super-extremal. Such states are generally thought to be non-physical. We will discuss extremality in more detail in \cref{sec:extremal}. The temperature of an RN black hole is given by
\begin{equation}
    \label{eq:rn-temperature}
    T = \frac{\left(M^2-Q^2\right)^{1/2}}{2\pi\left(M+\left(M^2-Q^2\right)^{1/2}\right)^2}.
\end{equation}

Hawking radiation has yet to be directly observed, due mainly to the fact that all known black holes have large masses, and are therefore extremely cold. An astrophysical black hole cannot form below the Chandrasekhar limit \citep{Chandrasekhar:1931ih} of $\sim1.4M_\odot$, for which the corresponding temperature is $T\sim\SI{4e-12}{\electronvolt}$. Thus, the effects of Hawking radiation on astrophysical black holes are negligible even on cosmological timescales. Since all known black holes are cold, with temperatures much lower than the masses of any known massive particles, black hole evaporation is often treated by considering only the emission of neutral massless particles. But in our scenario, we are interested in black holes of primordial origin, which may form with much lower masses, and thus radiate with much higher temperatures. Such black holes can produce massive charged particles at an appreciable rate.

Since there is no need for such particles to be emitted in pairs of opposite sign, a neutral black hole can spontaneously acquire an electric charge by emission of a charged particle. On the other hand, a charged black hole is more likely to emit particles of like sign \citep{Gibbons:1975}, so the spontaneous charge of a sufficiently small black hole fluctuates rapidly around neutrality. \citet{Page:1977um} studied the distribution of black hole charges numerically, and found that if a black hole is small enough to emit charged leptons rapidly, the equilibrium charge distribution is approximately Gaussian,
\begin{equation}
    \label{eq:page-distribution}
    P(Q) \sim \exp\left(-4\pi\alpha(Q/e)^2\right),
\end{equation}
with rms value of $Q/e$ given by $(8\pi\alpha)^{-1/2}\approx2.34$. The numerical calculations in that work show that if the product of the black hole mass and the emitted particle mass is small in Planck units, then the rms value of $Q/e$ increases to $\sim6$.

In our scenario, we envision that evaporation is halted near the Planck scale, and that any remaining charge is thus ``stuck'' on the black hole, leaving a charged Planck-scale relic (CPR). Of course, black hole evaporation is not well understood at masses near the Planck scale, and the outstanding issues in the study of black hole evaporation are beyond the scope of this work. Ultimately, we must neglect these problems in order to study the basic plausibility of our scenario. However, we will first review what the problems are, and discuss which ones can be ameliorated in our context and which ones cannot.

The spontaneous emission of charge by black holes has been studied analytically, e.g. by \citet{Gibbons:1975}, and one might hope that such analytical work could serve as a guide for our study. However, such analytical techniques break down when the black hole horizon becomes smaller than the emitted particle's Compton wavelength. Thus, we must retreat to numerical techniques. In the ultra-low-mass regime, near $\mpl$, there are several additional issues that confound an exact calculation of the charge distribution. Of course, the behavior of gravity itself is poorly understood in this regime: quantum gravity corrections should be significant, and it is not known how this influences the charge distribution. But even treating gravity as a classical background, several problems remain.

The first problem is the treatment of backreaction from emitted charges on the rate of subsequent emissions. The relevant quantity here is the timescale separating distinct emission events. For massive black holes, with low temperatures, this timescale is quite long \citep{Page:1977um}, and backreaction can be neglected. But for small black holes, the emission rate is much higher, so it may be inappropriate to treat consecutive emission events as independent processes. The nature of backreaction and its connection to black hole stabilization is subject to ongoing discussion in the literature \citep[see e.g.][]{Paul:2016xvb}, and the impact on the charge distribution is unclear.

The second problem is that as the mass becomes very small, the charge-to-mass ratio becomes appreciable, and the impact of the charge on the black hole geometry cannot be neglected. This is manifested most clearly in the case that $Q\sim12e$ for a black hole of $M\sim\mpl$, in which case the black hole is \emph{near-extremal:} the charge-to-mass ratio is nearly as great as possible, and the surface gravity of the black hole drops nearly to zero. An exactly extremal black hole has a temperature of exactly zero, and emits no thermal Hawking radiation. (It may still radiate athermally, as we will discuss shortly.) The calculation of \citet{Page:1977um} assumed that $Q/M\ll1$, a condition we may very well violate in our scenario.

The third problem concerns the role of the electromagnetic coupling $\alpha$. At large black hole masses, the width of the equilibrium charge distribution in \cref{eq:page-distribution} is sensitive to $\alpha$. The calculation is perturbative, so it is critical that the back-reaction of emitted particles on the metric should be higher-order in $\alpha$. But this is not necessarily the case at extremely small length scales. To make matters worse, the temperature is also of order the Planck scale, meaning that the relevant value of $\alpha$ is subject to renormalization all the way to the Planck scale, and thus is sensitive to potentially all of BSM physics.

In light of all these issues, it is impractical to attempt a first-principles calculation of the charge distribution of relic Planck-scale black holes. Thus, in this work, we only perform an extremely naive estimate of the charge fraction as a plausibility argument, and then outline how such massive charged objects could be detected.

\subsection{Emission from black holes}
\citet{Hawking:1975} showed that for a species with charge $q$ and mass $m$, the emission rate at a frequency $\omega$ in each angular mode $(\ell,m)$ and polarization $p$ is given by 
\begin{equation}
    \label{eq:species-emission-rate}
    \frac{\du N_{\ell,m,p}}{\du t\dd\omega} =
        \frac{\Gamma_{\ell,m,p}(\omega,T,q\Phi)/2\pi}
             {\exp\left[\left(\omega+q\Phi\right)/T\right]\pm 1}
\end{equation}
where $\Phi$ is the electrostatic potential at the surface of the hole  ($-Q/r_+$ in our case), and $\Gamma_{\ell,m,p}$ is an absorption coefficient specific to that mode. The emission rate of~\cref{eq:species-emission-rate} has the form of a thermal spectrum with a chemical potential proportional to the black hole's charge. It is sometimes useful to take a different viewpoint, and consider the emission rate to result from a combination of two mechanisms, one thermal and one athermal.

Heuristically, Hawking emission can be viewed as the separation of spontaneous virtual particle-antiparticle pairs by the black hole horizon. In the absence of charge, this process is mediated by gravity alone. This is the ``thermal'' component of black hole evaporation, which deviates from a blackbody spectrum only by virtue of the greybody factors $\Gamma_{\ell,m,p}$. However, if the black hole has a significant charge, then the picture must be modified: now, in addition to strong curvature near the horizon, there is a strong electric field. A strong electric field, even in the absence of curvature, can separate particle-antiparticle pairs in much the same way. This particle production due to vacuum polarization is just the familiar Schwinger mechanism \citep{Schwinger:1951}. It enters into \cref{eq:species-emission-rate} in two ways: first, as a chemical potential in the exponential factor, and second, via the dependence of $\Gamma_{\ell,m,p}$ on the black hole's charge. Note that \cref{eq:species-emission-rate} is compatible with the operation of the Schwinger mechanism even when the black hole's temperature is exactly zero. We refer to the component of radiation associated with the Schwinger mechanism as \emph{athermal} emission, and we refer to the remainder as \emph{thermal} emission. We will discuss the consequences of these two components further in \cref{sec:extremal}.

In a sense, these two mechanisms compete: thermal emissions drive the black hole away from neutrality in a random walk, but the athermal emissions are always of like sign to the black hole, and tend to discharge it. Equivalently, the black hole emits charges of both signs as long as $|Q|<M-e$, but as $|Q|$ increases, the emissions are increasingly biased to have the same sign as $Q$. This fact led \citet{Gibbons:1975} to observe that a small black hole cannot maintain even one elementary charge for an appreciable length of time, so long as evaporation remains active. We are interested in the characteristic lifetime of both neutral and charged black holes, where any charge implies a significant charge-to-mass ratio due to the small mass, so we cannot neglect either the thermal or the athermal component. Thus, it is important for us to compute the absorption coefficient $\Gamma_{\ell,m,p}$ for charged leptons to the extent possible in our framework. The absorption coefficient is calculated by solving the Dirac equation for an incoming wave with the appropriate boundary conditions in the Reissner-N\"ordstrom geometry \cite{Teukolsky:1974}. The solution can be resolved into ingoing and outgoing waves, from which transmission and absorption coefficients can be extracted. This has been done numerically for several particle species by \citet{Page:1976a,Page:1976b,Page:1977um}, resulting in the distribution of \cref{eq:page-distribution}.

In principle, this distribution applies even to black holes with large masses. However, the time to reach equilibrium grows with the timescale of lepton emission. For black holes whose Hawking temperatures are below the lowest lepton mass---and certainly for astrophysical black holes---this timescale is extremely long, and we should expect the charge distribution of such black holes to be dominated by accretion of charged particles instead of evaporation. On the other hand, in the low-mass regime, where the Hawking emission timescale is very short, any charge acquired due to accretion will quickly be erased by evaporation processes, and the equilibrium distribution will be maintained.

In light of the issues discussed in this section, it is inappropriate to directly extrapolate the charge distribution of \citet{Page:1977um} to the Planck scale. Instead, in an effort to account for as many low-mass effects as possible, we implement a similar numerical calculation, and extract an order-of-magnitude estimate of the fraction of black holes with non-zero charge. We describe this calculation in the following section.

\subsection{Estimating the charged fraction}
In light of the problems discussed in~\cref{sec:evaporation}, it is infeasible to perform a first-principles calculation of the fraction of stalled relics with spontaneous charge. However, we can perform a naive estimate by applying results developed for massive black holes, discarding approximations wherever possible. This result will not be a robust prediction of the charged fraction, but will instead represent a semi-classical guess. In this section, we make such an estimate, and then determine the implied abundance of CPRs today.

We can perform a first estimate of the charged fraction in the relic population by evaluating two timescales: the characteristic timescale $\tau_\mathrm{neutral}$ for a neutral black hole to acquire a non-zero spontaneous charge, and the characteristic timescale $\tau_\mathrm{charged}$ for a black hole with charge $Q=e$ to discharge and become neutral. Then the fraction of objects which are charged at the moment that evaporation stalls can be estimated as
\begin{equation}
    f_{\mathrm{charged}} = \left(1+\frac{\tau_\mathrm{neutral}}{\tau_\mathrm{charged}}\right)^{-1}.
\end{equation}
The timescale $\tau_{\mathrm{charged}}$ can be bounded below by neglecting time spent at higher charges during the black hole's semi-random walk, and evaluating only the minimum time to discharge, i.e., 
\begin{equation}
    \label{eq:lifetime-charged}
    \tau_{\mathrm{charged}} \gtrsim \left.e\middle/
        \left.\frac{\du Q_-}{\du t}\right|_{Q=e}
    \right..
\end{equation} 
Here we use the notation $\du Q_-/\du t$ to denote the rate of emission in \emph{positive} charge only, i.e., the rate at which the spontaneous charge decreases, as though by addition of negative charge. Likewise, $\du Q_+/\du t$ denotes the rate of emission in negative charge only. The overall evolution of the charge is governed by $\langle\du Q/\du t\rangle=\du Q_+/\du t - \du Q_-/\du t$ on timescales that are long compared to the emission rate. It is critical to distinguish between the signed emission rates and the average, since $\langle\du Q/\du t\rangle=0$ for $Q=0$, while $\du Q_\pm/\du t$ are individually non-zero. Since a neutral black hole can decay to a state with either sign with equal probability, we have $\du Q_+/\du t=\du Q_-/\du t\equiv\du Q_\pm/\du t$, and the lifetime of the neutral state can be estimated as 
\begin{equation}
    \label{eq:lifetime-neutral}
    \tau_{\mathrm{neutral}} \simeq \frac12\left.e\middle/
        \left.\frac{\du Q_\pm}{\du t}\right|_{Q=0}
    \right..
\end{equation}
Then the charged fraction can be estimated as 
\begin{equation}
    f_{\mathrm{charged}} \gtrsim \left(
        1 + \frac12 \frac{
          \left.\du Q_-/\du t\right|_{Q=e}
        }{\left.\du Q_\pm/\du t\right|_{Q=0}}
    \right)^{-1},
\end{equation}
so our task is to compute $\du Q_\pm/\du t$ for a black hole near the Planck scale, with a charge of either zero or $e$. We neglect higher charges since such black holes should neutralize more rapidly. The effect of including them would only be to increase the final charged fraction.

\citet{Page:1977um} evaluates $\du Q_-/\du t$ for massive black holes following \cref{eq:species-emission-rate}, computing the absorption probability $\Gamma_{\ell,m,p}$ numerically. The relevant information is found in fig. 4 of that work, which shows the emission rate ($\du N/\du t$) of charged leptons from a black hole as a function of the black hole charge for $-25e\leq Q\leq 25e$. \citet{Page:1977um} has calculated this rate separately for values of $M\mu$ in increments of 0.1 between 0.00 and 0.40 in Planck units, where $M$ is the black hole and $\mu$ is the mass of the emitted lepton. In our case, $M\sim1$ and $\mu\sim m_e/\mpl$, so $M\mu=0$ is the appropriate choice. Extrapolating those results to our regime, we find that
\begin{equation}
\frac{
          \left.\du Q_-/\du t\right|_{Q=e}
        }{\left.\du Q_\pm/\du t\right|_{Q=0}} \approx 1.02.
\end{equation}
In short, this indicates that the athermal Schwinger emissions are at most comparable in rate to thermal Hawking emissions. If this is indeed the case, and given that we have neglected higher charges, the charged fraction is $f_{\mathrm{charged}}\gtrsim1/2$.

\begin{figure}
    \includegraphics[width=0.49\textwidth]{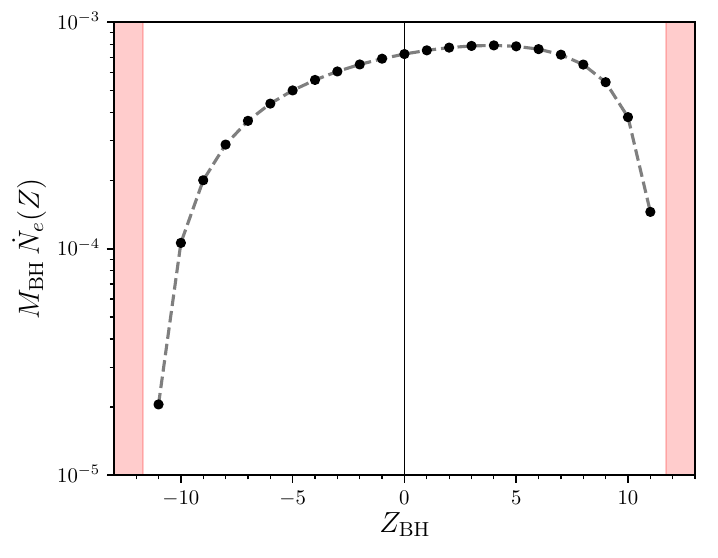}
    \hspace{0.02\textwidth}
    \includegraphics[width=0.49\textwidth]{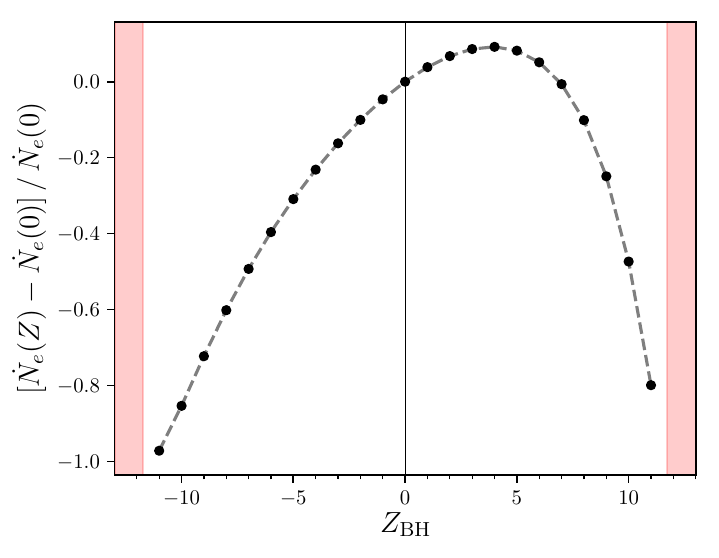}
    \caption{\textbf{Left:} the emission rate of (positive) particles of charge $+e$ from a black hole with charge $Ze$. Note that the charge number $Z$ is shown in electron units, not Planck units, so that an extremal black hole has $Z\approx11.7M/\mpl$. The mass of the black hole is fixed to $\mpl$ for the purposes of this calculation. This corresponds to the line ``$M\mu=0.00$'' in fig. 4 of \citet{Page:1977um}, but differs in that we carry out the computation for a mass which is not orders of magnitude larger than the charge. In particular, a substantial modification to the emission rate is observed when $|Q|\sim M$. The red shaded regions indicate where the black hole is super-extremal. \textbf{Right:} the fractional difference between the emission rate at $Q=Ze$ and the emission rate at $Q=0$, computed at $\mpl$. Note the linear scale. The asymmetry in both the left and right plots is due to the electrostatic potential, which behaves like a chemical potential, and enhances the rate of emissions which tend to neutralize the black hole. However, for $|Q|\sim e$, the emission rate is modified by only $\sim5\%$.}
    \label{fig:charge-rate}
\end{figure}

However, even insofar as we are neglecting the failure of various approximations in the limit $M\to\mpl$, the results of \citet{Page:1977um} cannot be directly applied. In that work, the numerical calculations themselves were always performed with $M\gg \mpl$. The label $M\mu=0.00$ does not suggest that the result applies to our case, in which $M\mu$ is vanishingly small: $M\mu\sim10^{-23}$. But even though we cannot assess the impact of Planck-scale physics on these results, we can still remove the uncertainty associated with the numerical computation by re-implementing the calculation and inserting this actual value of $M\mu$. Further, we solve eqs. (15) and (16) of \citet{Page:1977um} without neglecting the charge, as was done in that reference. The results are shown in~\cref{fig:charge-rate}. The implication is qualitatively unchanged for $Q\sim e$, with $\tau_{\mathrm{neutral}}/\tau_{\mathrm{charged}}\lesssim1.04$.

Generally speaking, as shown in fig. 4 of \citet{Page:1977um}, this ratio approaches unity as the mass of the black hole decreases. This is to be expected: the emission rate in same-sign particles scales with the absorption coefficient for modes that discharge the black hole, and in the low-mass limit, this coefficient is strongly suppressed \cite{Gibbons:1975}. The black hole still tends to discharge rapidly in the absence of charge fluctuations, but in our case, the timescale for discharge becomes comparable to the timescale of an upward charge fluctuation. This estimate of the charged fraction does assume that the evaporating black hole rarely enters the near-extremal regime ($Q\sim M$). However, as discussed in \cref{sec:extremal}, we expect near-extremal states to have a very short lifetime due to athermal emission. Further, the potential stalling of evaporation due to extremality makes no significant difference to the outcome: since the rms charge distribution has a width of order $1$--$10e$, and an extremal hole must have $Q/e\sim10M$, we do not expect extremality to be an important consideration unless $M\sim 1$ already. In any case, extremality effects can only increase the charged fraction.

None of this discussion overcomes the fundamental difficulties with calculations near the Planck scale. However, this calculation establishes that in the absence of some new physics or new phenomenology, we should generically expect at least $\sim50\%$ of black holes to be charged at the end of their evaporation. If their evaporation is halted, it is thus plausible that the relic population has a significant charged fraction.

\subsection{Near-extremal regime}
\label{sec:extremal}
As the black hole mass approaches $\mpl$, there is an additional (classical) complication: the charge-to-mass ratio approach unity, changing the geometry of the black hole significantly. If the black hole's charge undergoes $\order(e)$ fluctuations, then our scenario involves charge-to-mass ratios of at least $e/\mpl\approx 1/11.7$. If the charge fluctuates even briefly to $\order(10e)$, then we may expect to have $Q/M\sim 1$ at some point during the black hole's evolution. This is the \emph{near-extremal} regime. In this section, we review the properties of extremal Reissner-Nordstr\"om black holes and discuss the implications for the relic population.

An extremal black hole is a black hole with vanishing surface gravity, or equivalently, one whose mass is the smallest possible for its charge and angular momentum. The self-energy of the electric field and the angular momentum can both be thought of as contributing to the mass, so at fixed charge, the mass cannot be decreased arbitrarily. Super-extremal black holes, i.e., those with charge beyond the extremal limit, violate the cosmic censorship conjecture and are generally considered unphysical. In the Reissner-Nordstr\"om case, an extremal black hole has $Q/M=1$, meaning that $r_+=r_-$. This corresponds to a charge of $1/\sqrt\alpha\approx 11.7e$ for a black hole of mass $\mpl$.

Since the temperature of a black hole is proportional to its surface gravity, an \emph{exactly} extremal black hole does not produce any \emph{thermal} Hawking radiation. Indeed, this is required by the cosmic censorship conjecture: any neutral emission would reduce the mass of the black hole without a commensurate reduction in charge, leading the black hole to be super-extremal. The temperature decreases smoothly in the near-extremal regime (see \cref{eq:rn-temperature}). At first glance, this seems to provide a potential mechanism for the stability of CPRs, completely apart from Planck-scale physics: if $T=0$, it is tempting to conclude that the black hole does not radiate. In this case, we could suppose the charge of the black hole originally fluctuates rapidly, following a distribution with some width. Then, as the mass decreases, the black hole might become extremal, stalling the evaporation process. There are two problems with this idea: first, extremal black holes are not necessarily stable, even though the Hawking temperature vanishes. Secondly, even if a near-extremal geometry stabilizes the black hole, this can only manifest if the black hole is extraordinarily close to extremality. Note that we only consider near-extremal states due to the third law of black hole thermodynamics, which states that a black hole cannot evolve to an exactly-extremal state. We will discuss each of these issues in turn.

Regarding stability, the thermodynamics of near-extremal black holes is still not fully understood. They certainly cannot emit neutral particles if the cosmic censorship conjecture holds. But all known charged particles satisfy $|q|>m$, so it may be possible for an extremal black hole to athermally emit charged particles. If not, then there exists an infinite set of stable extremal states, one for each charge number. The existence of such an infinite tower of stable states without any corresponding gauge symmetry is believed to be incompatible with string-theoretic UV completions \citep{Vafa:2005ui}. This is a major motivation for the weak gravity conjecture \citep[WGC,][]{ArkaniHamed:2006dz}, which requires that for \emph{any} gauge symmetry, there exists a state with charge $|q|>m$. The WGC was posited in part to allow extremal black holes to be unstable, precisely in order to prevent the appearance of such an infinite tower in the spectrum of a theory.

Note that athermal emission in the extremal state is consistent with the emission rate in \cref{eq:species-emission-rate}: in the limit of small $T$, the rate vanishes if the argument of the exponential is positive. On the other hand, for a sufficiently large and negative chemical potential $q\Phi$, the argument of the exponential is negative, and the low-temperature limit is proportional to $\Gamma_{\ell,m,p}(\omega,T,q\Phi)$. Recall that $\Phi=-Q/r_+$ in our case, and for an extremal RN black hole, $r_+=M=Q$. Then the argument of the exponential is negative as long as $q\Phi<-\omega$, so the requirement for the black hole to produce a particle of charge $q$ and mass $m$ is exactly that $|q|>m$. Thus, resolving the question of stability depends on the calculation of the coefficient $\Gamma_{\ell,m,p}$. Heuristically, the black hole could still radiate because the electric field at the surface of the black hole remains strong, so particle-antiparticle pairs could still be separated by the Schwinger mechanism, even though the surface gravity vanishes. Several authors \citep{Vagenas:2000am,Khriplovich:2002qn,Chen:2014yfa} discuss emissions from extremal and near-extremal states in more detail.

Regarding the physicality of the extremal state, the third law of black hole thermodynamics is analogous to the ordinary third law of thermodynamics, which implies that no statistical system can attain a temperature of exactly zero (see e.g. \citet{Belgiorno:2004} for an extensive discussion). The applicability of statistical laws to small black holes remains an active area of research \citep[see e.g.][]{Preskill:1991tb}, but the situation is readily understood heuristically: as a black hole approaches extremality, its temperature decreases according to \cref{eq:rn-temperature}, and its emission rate decreases. This may stall evaporation temporarily, but the black hole cannot evaporate to an extremal state by this mechanism. The next question, then, is whether such stalling is significant on cosmological timescales. The stalling of evaporation near extremality was studied numerically for large black holes by \citet{Hiscock:1990}. They conclude that although the third law is satisfied at all times, the reduction of the emission rate in the near-extremal limit can prolong the black hole lifetime considerably.

However, our scenario is substantially different in that the evaporation cannot be treated smoothly: since we are interested in a phase of black hole evolution which involves extremely high temperatures, the black hole can lose charge by emission of charged leptons on timescales that may be relevant to mass loss. The appropriate analogue of the analysis of \citet{Hiscock:1990} would be to solve a system of differential equations for the evolution of the joint probability distribution of mass and charge, $\mathcal P(M,Q)$, treating charge as discrete---that is, a system of the form
\begin{multline}
    \frac{\du\mathcal{P}(M,Q)}{\du t}=-\sum_{Q^\prime\in\mathbb Ze}\int_0^{M}\du M^\prime\;\mathcal R(M\to M^\prime;\;Q\to Q^\prime)+\\
    \sum_{Q^\prime\in\mathbb Ze}\int_{M}^\infty\du M^\prime\,{\mathcal P}(M^\prime,Q^\prime)\mathcal R(M^\prime\to M;\;Q^\prime\to Q),
\end{multline}
where $\mathcal R(M\to M^\prime;\;Q\to Q^\prime)$ gives the differential rate for black hole of mass $M$ and charge $Q$ to decay to a black hole of mass $M^\prime$ and charge $Q^\prime$. Further, while \citet{Hiscock:1990} use an approximate form of the emission rate which is valid only for sufficiently massive black holes, $\mathcal R$ must be the full rate in our case, as computed from \cref{eq:species-emission-rate}. Under these conditions, this system is difficult to solve numerically, especially since any numerical evaluation must be sensitive to extremely small values of $M-Q$.

Thus, a simpler estimate of the near-extremal behavior is called for. First, observe that in our case,  in order for the charge to stabilize, we must at least have $T\lesssim m_e$, or else thermal emissions alone will cause charge fluctuations. But in our scenario, black holes only have an appreciable probability of being within $1e$ of extremality when $M\sim\mpl$---and even then, the probability is $\order(1\%)$ if we naively extrapolate the distribution of \cref{eq:page-distribution}. For a black hole with $M\sim\mpl$, to have a temperature of $T < m_e$, the charge-to-mass ratio must be extremely close to unity. For fixed $Q$, the mass $\mfreeze$ at which thermal electron production is frozen out ($T = m_e$) is given by
\begin{equation}
    \mfreeze(Q) = Q + 2\pi^2 Q^3 m_e^2 + \mathcal O(m_e^3),
\end{equation}
where we recall that, in Planck units, $m_e\ll 1$. If $Q=12e$, then we must have $\mfreeze - Q \simeq 4\times10^{-44}$, so the hole's mass must depart from its extremal value by no more than $\delta M\simeq \SI{5e-16}{\electronvolt}$. This is an extremely small ``target'' to hit: outside of this region, the power of emission in charged particles is comparable to that in neutral particles, so the charge is likely to fluctuate on the same timescale that governs the shrinking of the mass.

This alone does not make it impossible for the hole to enter the near-extremal regime, but it does make this unlikely to take place during a typical evaporation. To estimate the probability, we give the following argument: the hole is most likely to be near-extremal when the mass is lowest. Thus, suppose that the hole passes through the mass range $12e<M<\mfreeze(12e)$ at some time. What is the probability that the charge takes the value $12e$ at some moment during this interval, i.e., before the hole evaporates further to $M<12e$? Since $\mfreeze(12e)-12e<m_e$, the black hole charge cannot change without the mass dropping to $M<12e$. This means that the black hole must already have charge $12e$ when $M=\mfreeze(12e)$. If the spontaneous charge distribution for such a small hole is at all similar to that of its larger counterparts, which have rms charges of order $6e$, this situation is quite unlikely, happening with a probability of a few percent.

However, there are three scenarios in which extremality may be significant. First, suppose that evaporation of a neutral black hole does not stall, or that it stalls at a scale $\mstop\ll\mpl$, as is possible in the context of the generalized uncertainty principle \citep{Carr:2015nqa}. In this case, the black hole may enter a mass regime where the maximal charge is comparable to or smaller than the width of the spontaneous charge distribution, and the chance of freezing out charged leptons can no longer be neglected. Note that in the sub-Planckian regime, evaporation can behave very differently. For instance, in the case of \citet{Carr:2015nqa}, $T\sim M$ for $M<\mpl$, which may substantially modify the results of our analysis. Second, if a sufficiently large population of near-Planck-scale black holes is produced in the early universe, and only near-extremal holes are stable, then even a tiny fraction of this initial population could account for a significant fraction of dark matter. The hot evaporation products of the remainder would redshift away like radiation. Third, any charge associated with a new $\mathrm U(1)$ symmetry would influence the metric in the same way as electric charge. In particular, the fine structure constant associated with the new symmetry could be much smaller than $\alpha_{\mathrm{EM}}$, smoothing out the discrete spontaneous charge distribution. Alternatively, the lightest charged state of the new symmetry could be much more massive than the electron, reducing its athermal production rate, and making near-extremal states long-lived \citep{Bai:2019zcd}.

For the remainder of this work, we will not need to assume that CPRs are extremal. However, we note that if even a small fraction of the initial PBH population does evolve to a state sufficiently near extremality to freeze out thermal lepton emissions, and if this state is stable to athermal emissions as well, then this makes the CPR scenario viable even if PBH evaporate completely. In this case, the dark matter density is fixed by the initial number of evaporating PBH and the fraction that freeze, and all surviving PBH are near-extremal. However, we note that if extremality is the only stabilizing mechanism, and if these objects accrete opposite charge at any time, then they are unlikely to stabilize again into a charged state. Instead, they will completely evaporate. If such destabilization events are still ongoing in the late universe, this will result in potentially observable bursts of high-energy particles.

\subsection{Cosmic history of CPRs}
\label{sec:history}
The existence of CPRs today requires a primordial origin for the original generation of black holes. In this section, we examine the feasibility of producing a detectable population of CPRs through such a mechanism, starting with their formation in the early universe.

In the simplest scenario, the progenitors of CPRs are produced near the Planck scale with a monochromatic mass function. However, black holes need not be dominantly produced near the Planck scale in order to leave behind CPRs today. Multi-modal mass functions have been invoked to account for all of dark matter while avoiding constraints. More generally, primordial black holes can be produced with an extended mass function, e.g. with a lognormal or power-law mass function \citep{Carr:2016drx}, and such a broad initial spectrum will typically produce a small abundance of relics by evaporation. Any primordial black hole produced with a mass below $M_{\mathrm{evap}}\sim\SI{5e14}{\gram}$ will evaporate to the Planck scale by the present day \citep{Page:1977um}. Further, any black hole produced with a mass below $\sim\SI{e16}{\gram}$ has an initial temperature of order \SI{1}{\mega\electronvolt}, and thus produces charged leptons rapidly enough to acquire a spontaneous charge, even though it will continue to evaporate actively today.  As such, even if they are dominantly produced at even higher mass scales, we generically expect to find a low-mass tail that evaporates to the Planck scale---and might leave behind CPRs.

There are two major constraints on such a scenario: first, even if a relic is left behind at the end, the total radiation produced by evaporating black holes is constrained by CMB observables and light element ratios \citep{Carr:2009jm}. Second, if the CPRs originate from a population of black holes with a component above $M_{\mathrm{evap}}$, then they do not constitute all of dark matter, and may indeed account only for a small fraction. In this case, other probes can constrain the population at higher masses. To investigate the plausibility of such a scenario, we suppose that all dark matter is in the form of PBHs, and suppose that the black holes above $M_{\mathrm{evap}}$ do not lose a significant amount of their mass. Then, given the initial mass function $\du n/\du M$, the density of CPRs today is given by
\begin{equation}
    \frac{\Omega_{\mathrm{CPR}}}{\Omega_{\mathrm{DM}}} \approx \frac{
        \mpl\int_{\mpl}^{M_{\mathrm{evap}}}\du M\,\frac{\du n}{\du M}
    }{
        \mpl\int_{\mpl}^{M_{\mathrm{evap}}}\du M\,\frac{\du n}{\du M}
        + \int_{M_{\mathrm{evap}}}^\infty \du M\, M\frac{\du n}{\du M}
    }.
\end{equation}
For example, as a toy model, consider a power-law mass function $M \dd N/\du M\propto M^{\gamma-1}$. Assuming $\gamma<0$, the mass fraction in CPRs is 
\begin{equation}
    \frac{\Omega_{\mathrm{CPR}}}{\Omega_{\mathrm{DM}}} \approx \frac{
        \mpl M_{\mathrm{evap}}^{\gamma-1} - \mpl^{\gamma}
    }{
        M_{\mathrm{evap}}^{\gamma-1}\left[
            \mpl - (1-1/\gamma)M_{\mathrm{evap}}
        \right] - \mpl^\gamma
    }.
\end{equation}
A CPR fraction $f\sim1$ is produced when $\gamma\lesssim-0.1$, whereas e.g. $f\sim10^{-2}$ for $\gamma\sim-10^{-2}$.

Next, we must consider the survival of such charged objects over cosmic time. It is unlikely that a Planck-scale black hole would neutralize by accretion of charged particles, since the geometric cross section is extremely small. In other words, we expect the accretion rate to be suppressed by $\mpl^2$. But even if a CPR does accrete, the consequent increase in mass may restart the evaporation process, and in the low-mass regime, the emission power in charged particles is comparable to that in neutral particles. If we treat accretion as an excitation of the black hole remnant to a neutral state with a higher mass, this excited state can simply decay again to a charged state. Thus, we expect neutralization of the CPR population to take place very slowly if it happens at all.

A more plausible scenario is that the black hole forms bound states with particles of opposite sign. For positively-charged CPRs, electron capture would take place alongside the same process for hydrogen atoms during the epoch of recombination. For negatively-charged CPRs, capture of protons is even more energetically favorable, since the energy of the bound state scales with the reduced mass.
At first glance, this could interfere with detection: a net-neutral bound state may be invisible to a terrestrial detector. We will show in \cref{sec:neutral-detection} that such objects are still detectable. But it is still important to understand the typical charge state of CPRs far from Earth, in part because accretion of a bound charge might be possible on cosmological timescales. Thus, we now examine their ionization history.

It is easily checked that reionization proceeds almost identically for CPRs as for hydrogen, even for negatively-charged CPRs with bound protons. In this case, following \citet{Madau:1998cd}, the CPR population will be fully ionized when emission rate of ionizing photons per unit volume matches the rate of recombinations, that is,
\begin{equation}
\label{eq:ionization-minimum}
    \dot n_\gamma \gtrsim \frac{n_{\mathrm{CPR}}}{t_{\mathrm{rec}}},
\end{equation}
where $n_{\mathrm{CPR}}$ is the number density of CPRs and $t_{\mathrm{rec}}$ is the characteristic timescale for recombination with a free proton. We can estimate this timescale as $t_{\mathrm{rec}}\simeq1/(n_{\mathrm{p}}\alpha_A)$, where $\alpha_A$ is the recombination coefficient. We calculate the recombination coefficient $\alpha_A$ for a bound proton following \citet{Boardman:1964}\footnote{
    Note that \citet{Boardman:1964} contains two typographical errors in eqs. (2) and (3). Correct versions of these equations can be found in \citet{Karzas:1960}.
}. At a typical nebular temperature of $\SI{e4}{\kelvin}$, we find that $\alpha_A\simeq\SI{2e-21}{\centi\meter^3/\second}$, versus $\alpha_A\simeq\SI{4e-13}{\centi\meter^3/\second}$ for hydrogen.

The much smaller recombination coefficient and number density imply that reionization should proceed much more rapidly for CPR atoms than for hydrogen: if CPRs at the Planck mass constitute all of dark matter, then their number density is still lower by a factor of $\sim10^{-19}$ compared to that of protons, and $t_{\mathrm{rec}}$ is reduced by $\sim10^{-8}$ compared to hydrogen. We can also account for the fact that the binding energy of a proton with a negatively-charged CPR is $\sim\SI{25}{\kilo\electronvolt}$ in the ground state, so the only ionizing photons are those with wavelength $\lambda\lesssim\SI{0.5}{\angstrom}$, compared with $\lambda<\SI{911}{\angstrom}$ for hydrogen. If we extrapolate the quasar spectrum of \citet{Madau:1998cd} to small wavelengths, where $L(\lambda)\sim\lambda^{1.8}$, the emission rate is suppressed by a factor of $\sim10^{-6}$ compared with photons that ionize hydrogen. This suppression is insignificant compared to the changes in the number density and recombination timescale, so we conclude that CPRs will still reionize much more efficiently than hydrogen.

\section{Detecting charged black holes terrestrially}
\label{sec:detection}
In this section, we analyze the interaction of CPRs with matter, and investigate mechanisms for direct detection. For the purposes of our calculations, we assume that CPRs account for a fraction $\fcpr$ of dark matter by mass (i.e., $\fcpr=\Omega_{\mathrm{CPR}}/\Omega_{\mathrm{DM}}$). We assume that evaporation is halted at a mass $\mstop$ which we allow to differ from $\mpl$, and we assume that $\mstop$ is independent of the black hole charge. However, we require that $\mstop\geq e\approx\mpl/11.7$, since a (classical) black hole with a mass below this threshold cannot have even one elementary charge without being super-extremal, which we prohibit. Note that in some models, the relic mass is far below the Planck scale, e.g. as in \citet{Carr:2015nqa}. Such relics may yet evade the constraints we set here.

Even when $\fcpr=1$ and $\mstop$ is minimal, direct detection of such charged CPRs is limited primarily by the flux of these objects: at such high masses, the number density of CPRs is much lower than that of typical particle dark matter candidates. The flux is $\Phi_{\mathrm{CPR}}\simeq (\rho_{\mathrm{DM}}/\mstop)v_{\mathrm{DM}}$, so taking $v_{\mathrm{DM}}=\SI{300}{\kilo\meter/\second}$ and $\rho_{\mathrm{DM}}=\SI{0.3}{\giga\electronvolt/\centi\meter^3}$ gives the event rate as
\begin{equation}
    \label{eq:flux-estimate}
    N = \SI{0.23}{\per\year}\fcpr\left(
        \frac{\mstop}{\mpl}
    \right)^{-1}\left(
        \frac{A_{\mathrm{detector}}}{\SI{1}{\meter^2}}
    \right)\mathcal E_{\mathrm{detector}}
\end{equation}
where $\mathcal E_{\mathrm{detector}}$ is the fraction of CPRs that will register an event in the detector. Since we are considering electrically-charged objects, there are detectors for which $\mathcal E_{\mathrm{detector}}\sim1$, as we will detail in the following subsection. However, for CPRs, we expect to have $\mstop/\mpl\sim1$. Thus, in order to achieve a detection rate of $\SI{1}{\per\year}$, a detector must have $A_{\mathrm{detector}}\gtrsim\SI{4.3}{\meter^2}$.

It is clear from this calculation that a typical dark matter direct detection experiment is unlikely to encounter more than one such object during its operational lifetime. However, as we will discuss in the next subsection, the passage of even one CPR through a detector has the potential to produce an extremely clear signature.

\subsection{Signatures of CPR transits}
In this section, we discuss the interactions of CPRs with particle detectors. The interactions of a CPR with matter are similar to those of slow-moving heavy ions. A CPR with $\mstop\sim\mpl$ does not slow down appreciably during its transit through a detector: its kinetic energy is $\sim\frac12\mpl(\SI{300}{\kilo\meter/\second})^2\approx\SI{6e21}{\electronvolt}$. This is to be compared to atomic binding energies, which are typically of order $\SI{1}{\electronvolt}$. Indeed, a CPR is so massive that an object with a downward trajectory will \emph{gain} $\sim\SI{130}{\electronvolt/\angstrom}$ from gravitational acceleration. Deflection is also negligible, even in a strong electromagnetic field, so a CPR will deposit energy along a very straight track. In this respect, a CPR transit can be distinguished from any standard background: energy will be deposited at a constant density at a low speed ($\sim\SI{0.3}{\meter/\micro\second}$) along a straight track.

While detection prospects for CPRs in any given experiment must ultimately be studied with more detailed modeling, we can still use general methods to estimate signatures of a CPR transit. As a CPR transits through a detector, it loses energy via Coulomb interactions with the target electrons and nuclei. The transferred energy may be detectable in the form of heat, ionization, or scintillation. Each of these signatures scales with the energy deposited during the transit, typically expressed in terms of the ``stopping power'', that is, the energy loss of the incident particle per unit distance traversed through the material. We will identify the stopping power, ionization yield, and scintillation yield for particular experimental configurations using numerical simulations, but we begin with a simple estimate of the stopping power.

Regardless of whether they constitute a significant fraction of dark matter, CPRs should be highly non-relativistic ($\beta\sim10^{-3}$). Since this is slower than the outer electrons of the target atoms, the calculation of stopping power in this regime differs greatly from the relativistic regime. In particular, the characteristics of CPR interactions with matter are similar to those of heavy ions. The stopping power for $\beta<0.05$ is well-described by Lindhard-Scharff-Schiott theory \citep{Lindhard:1963}, in which it is linear in the velocity and charge of the incident particle, and has no dependence on its mass. Thus, we can estimate the stopping power in our scenario by comparing to empirical results for the stopping of non-relativistic muons. In copper, the stopping power per unit target density for incident muons with $\beta=10^{-3}$ has been measured as $(\du E/\du x)/\rho_{\mathrm{target}}\sim\SI{30}{\mega\electronvolt/(\gram/\centi\meter^{2})}$ \citep[fig. 23.1 of][]{RPP2018}. For our purposes, it is useful to quantify the stopping power in transit through Earth and in semiconductor detectors, and silicon is a representative material for both. The linear stopping power in silicon is then
\begin{equation}
    \label{eq:linear-stopping-estimate}
    \frac{\du E}{\du x}\sim\SI{70}{\mega\electronvolt/\centi\meter}.
\end{equation}
Note that since we may have $\left\langle\left| Z_{\mathrm{CPR}}\right|\right\rangle > 1$, this is a conservative estimate.

This stopping power is likely too small to register in a typical calorimeter. Thus, if a CPR is to be detectable, it must produce an ionization or scintillation signature. Given the stopping power, the ionization yield depends again on the particle velocity and material properties. The stopping power has two components, corresponding to interactions with electrons (electronic stopping power) and with nuclei (nuclear stopping power). 
For highly relativistic particles, nuclear stopping is typically negligible compared to electronic stopping, but this is not the case in the highly non-relativistic regime. Indeed, the maximum energy that a CPR can transfer to a recoiling electron with mass $m_e$ is given by
\begin{equation}
    \Delta E_{\mathrm{max}} = 2m_e v_{\mathrm{DM}}^2 \approx \SI{1}{\electronvolt}.
\end{equation}
This maximum energy transfer is smaller than typical ionization energies, so direct ionization via electronic interactions is not likely to be efficient. On the other hand, the maximum energy transfer in a recoil with a nucleus of mass number $A$ is
\begin{equation}
    \Delta E_{\mathrm{max}} = 2m_A v^2_{\mathrm{DM}} = \SI{186}{\kilo\electronvolt}\times \left(\frac{A}{100}\right)\left(\frac{v_\mathrm{DM}}{\SI{300}{\kilo\meter/\second}}\right)^2,
\end{equation}
much higher than that of electronic recoils. Thus, we expect interactions between the CPR and nuclei to dominate in a typical detector.

While ionization and scintillation are most efficiently produced by electronic interactions, nuclear stopping can also produce these signals, since the recoil energy of the nucleus can be partially transferred to bound electrons. The attendant loss of efficiency, or \emph{quenching}, is expressed via the ratio of yields from nuclear and electronic scattering. Such \emph{quenching factors} are dependent on the target material, and values are typically measured experimentally \citep[see e.g.][]{Tretyak:2009sr,Mu:2013pja}.
To estimate the stopping power, ionization yield, and scintillation yield due to nuclear recoils, we used the Monte Carlo code \SRIM~\citep{Ziegler:2010}, which simulates the passage of ions through matter. We simulated ``hydrogen'' ions with \SRIM's maximum allowable particle mass of \SI{10000}{\amu}, where \SI{}{\amu} is the atomic mass unit, and with a velocity of \SI{300}{\kilo\meter/\second}. This corresponds to a kinetic energy of \SI{4.7}{\mega\electronvolt}, still very large compared to the binding energies relevant for the interaction. In any case, we performed our simulations in 1 micron-thick layers of detector material, so the change in the momenta of the simulated ions was negligible. In the following section, we discuss the results of our simulations and the implications for different experimental modalities.

\subsection{Detection mechanisms}
\label{sec:detection-mechanisms}
Here we briefly survey several detector technologies to evaluate whether they would be suitable for detecting CPRs.

\detectortype{Bubble chambers}
\citet{Hawking:1971} noted that charged Planck-mass black holes would leave tracks in bubble chambers, and speculated that unidentified tracks in previous experiments could be explained by the presence of these objects. A bubble will form in superheated fluid if the energy deposited within a critical radius exceeds a given threshold energy. For concreteness, we consider the response of the PICO experiment \citep{Amole:2016pye,Amole:2017dex}, whose bubble chamber has a threshold energy of \SI{3.3}{\kilo\electronvolt} for a critical radius $r_c = \SI{2e-8}{\meter}$.

From the \SRIM~output, we integrated the energy deposited in a sliding window of width $r_c$, and found that the deposited energy was sufficient to form a track with a linear bubble density of $\sim\SI{e5}{\per\meter}$. This is not surprising: PICO is highly sensitive to $\alpha$ decays, which generate nuclear recoils of similar energy to those from CPRs. Further, a straight bubble track would not be expected from weakly interacting massive particles (WIMPs), which are expected to only interact once in the detector volume, and would be distinct from background signals from neutrons, which leave jagged tracks. However, even large bubble chambers have insufficient area to place strong constraints on the flux of CPRs. The proposed 500L version of PICO \citep{PICO500L} would require several decades of continuous exposure to place any constraint on the abundance of CPRs.

\detectortype{Atmospheric fluorescence detectors}
Ultra-high energy cosmic rays incident on the atmosphere generate hadronic and electromagnetic showers which ionize nitrogen molecules that subsequently fluoresce, emitting visible light. Arrays of photomultiplier tubes, such as the High Resolution Fly's Eye (HiRes) observatory \citep{Abbasi:2004nz}, are capable of detecting this fluorescence and reconstructing the track of the cosmic ray. Since HiRes can detect emissions over an area of order \SI{1}{\kilo\meter^2}, this seems like an attractive way to detect a particle with a very low flux.

We evaluated the potential of atmospheric fluorescence detectors to observe the energy deposition from the passage of a CPR. In dry air at sea level, our \SRIM~calculation yielded an energy deposition of $\sim\SI{12}{\mega\electronvolt/\meter}$. Assuming an average of \SI{3.5}{\electronvolt/photon} and a (generous) fluorescence efficiency of 5\% \citep{Sokolsky:1989}, the photon yield is about \SI{5e4}{\per\micro\second} for a relic moving at \SI{300}{\kilo\meter/\second}.
At the surface, background light from stars, light pollution, and other sources is about \SI{5e5}{\meter^{-2}.\steradian^{-1}.\micro\second^{-1}}. Given that a typical photomultiplier tube in HiRes observes 1 square degree with a \SI{5}{\meter^2} mirror, the background event rate is $\sim\SI{100}{\per\micro\second}$.
A CPR must then pass within a few meters of the mirror for the signal to overcome background photons, reducing the effective area of this class of detectors considerably. Atmospheric detectors are thus unlikely to place strong constraints on CPRs even in a decade of operating time.

\detectortype{Cherenkov detectors}
An attractive possibility is to search for CPRs with neutrino detectors (e.g. IceCube \citep{Collaboration:2011ym}, Super-Kamiokande \citep{Fukuda:2002uc}) or imaging atmospheric Cherenkov detectors (e.g. VERITAS \citep{Weekes:2001pd}, HAWC \citep{DeYoung:2012mj}), since they have extremely large ($\sim\SI{}{\kilo\meter^2}$) effective areas. However, regardless of their origin, we expect CPRs to be highly non-relativistic. Thus, we do not expect any Cherenkov radiation to be emitted as they traverse these detector media. Instead, light would be produced only by scintillation and ionization processes. Such a signal is distinguishable from those produced by relativistic particles in that light would be emitted isotropically from the track rather than in the cone shape characteristic of Cherenkov light. But a CPR transit would be extremely slow, taking place on the order of several \SI{}{\milli\second}, compared to typical targets observed over a duration of order \SI{}{\micro\second}. Thus, even if the light from ionization is observable, detecting it would require a non-trivial triggering mechanism.

\detectortype{Dark matter searches}
CPRs are highly penetrating and ionizing, so a CPR transit would leave a distinct signal in semiconductor and liquid xenon detectors, including dark matter direct detection experiments and neutrinoless double beta decay searches \citep{Garfagnini:2014nes}. Since CPRs are negligibly slowed by their interactions with materials, a CPR would produce a straight track in a detector with a transit time of order \SI{1}{\micro\second} in a typical experiment, a unique signature. Further, while the thick layer of earth (overburden) covering these experiments significantly reduces background, it does not affect the flux of CPRs. However, the flux itself is small, and these detectors typically have small cross-sectional areas. Even next-generation experiments are unlikely to detect more than a few CPRs in ten years, so they cannot produce significant constraints on the CPR population.

\detectortype{Monopole searches}
Magnetic monopoles are expected to be found at very large masses, and due to their high magnetic charges, monopole transits share some characteristics with CPR transits. As such, it is possible that monopole searches can impose constraints on CPRs as well. In \cref{sec:macro}, we investigate this possibility in detail in the context of the MACRO experiment \citep{Ambrosio:2002qq}.

\detectortype{Liquid argon detectors}
Liquid argon time-projection chambers \citep{Rubbia:1977zz} have recently been employed in several neutrino experiments \citep{Rubbia:2011ft,Anderson:2012vc,Cavanna:2014iqa,Antonello:2015lea,Abi:2017aow,Acciarri:2016ooe,Acciarri:2016smi}, some of which have much larger cross-sectional area than is typical for dark matter experiments. Since the transit of a CPR has a unique signature, backgrounds are of no concern, so existing neutrino experiments have the potential to detect CPRs. We expect that liquid argon detectors can be used to place strong constraints on the CPR population, and we elaborate on detection prospects in the ICARUS experiment in \cref{sec:icarus}.

\detectortype{Paleo-detectors} Recently, \citet{Baum:2018tfw} proposed the detection of WIMP dark matter through small tracks left in ancient minerals by dark matter recoils \cite[see also][]{Drukier:2018pdy,Edwards:2019puy}. The major strength of such ``paleo-detectors'' is their exposure time, of order \SI{e9}{\year}. This is uniquely well-suited to our case of interest, where detection is primarily limited by flux rather than detector efficiency. Moreover, we need not await the results of WIMP searches in ancient minerals: the recent paleo-detector proposals are extensions of similar searches already performed e.g. by \citet{Ghosh:1990} to constrain the flux of supermassive magnetic monopoles. These results should already be applicable to CPRs, and we discuss them in \cref{sec:paleo-detectors}.

\subsection{Detection of CPR ``atoms''}
\label{sec:neutral-detection}
In \cref{sec:history}, we argued that if CPRs form net-neutral bound states with electrons or atomic nuclei, they should be fully reionized by the present day. This applies to the astrophysical population of CPRs, but not necessarily to the terrestrial population relevant for direct detection. While enroute to a detector, an originally bare CPR may recombine with electrons or atomic nuclei to form a bound CPR ``atom'' (or ``neutraCHAMP'', in the language of \citet{Dimopoulos:1989hk}).

First we consider the possibility of recombination in the atmosphere, following \citet{Dimopoulos:1989hk}. In this case, positively-charged CPR with bound electrons will be rapidly ionized by solar UV radiation. The probability of recombination with an atmospheric electron is given by
\begin{equation}
    P_{\mathrm{rec}} = \sigma_{\mathrm{rec}} \rho_{e^-} L,
\end{equation}
where $\sigma_{\mathrm{rec}}$ is the recombination cross section,  $\rho_{e^-}$ is the free electron density, and $L$ is the depth of the atmosphere. We take $\sigma_{\mathrm{rec}}=10^{-5} \pi a_0^2$, $L=\SI{100}{\kilo\meter}$, and conservatively estimate $\rho_{e^-} = \SI{e6}{\centi\meter^{-3}}$, which yields $P_{\mathrm{rec}}\approx 10^{-3}$. Negatively-charged CPRs bound to a nucleus of atomic number $A$ will not be ionized by solar UV, as the binding energy is $\sim Z^2\left(\SI{25}{\kilo\electronvolt}\right)$. In this case, \citet{Dimopoulos:1989hk} calculated the mean free path for recombination with a \ce{^{14}N} nucleus to be $\lambda_{\mathrm{rec}} = \SI{4e10}{\gram.\centi\meter^{-2}}\beta^2/\rho_{\mathrm{atm}}\gtrsim\SI{300}{\kilo\meter}$, corresponding to a recombination cross section of $\sigma_{\mathrm{rec}}\simeq\SI{6e-34}{\centi\meter^2}$.

However, it is not clear that the latter cross section is applicable to interactions with atmospheric gas. Experimentally, cross sections for charge transfer onto slow ions in gaseous \ce{CO} and \ce{CO2} are much larger, of order \SI{e-16}{\centi\meter^2} \citep{Werbowy:2016}. The corresponding mean free path in the atmosphere is microscopic. Additionally, in passing through solid earth (overburden) enroute to a detector, a bare CPR would be very likely to acquire bound charge: experiments and simulations show that the recombination timescale of molecular hydrogen ions in carbon is $\order(\SI{10}{\femto\second})$ \citep{Cue:1980,Susuki:1994,Garcia-Molina:2003}. Thus, we must consider the possibility of detecting CPRs bound into neutral ``atoms''.

The stopping of partially-ionized or neutral atoms in materials differs from that of bare ions in that the screening effect of bound charges substantially modifies the potential. This generally diminishes but does not eliminate the stopping power. Numerous efforts have been made to model the stopping of neutral atoms \citep{Wilson:1977,Brandt:1982,Arnau:1990,Grande:1993,Sigmund:1997,Gobet:2006}, and much work in particular has been devoted to the development of an ``effective charge'' for such objects. In many circumstances, the effective charge of a neutral atom of atomic number $Z_1$ is $Z_{\mathrm{eff}}\simeq 0.7Z_1$ \citep{Brandt:1982}. However, the effective charge is generally a function of parameters beyond the nuclear charge and ionization state, including the atomic number $Z_2$ of the target and the velocity of the projectile \citep{Sigmund:1997}.

We are interested in the low-velocity, low-$Z_1$ regime for a wide variety of target materials---from xenon ($Z_2=54$) to argon ($Z_2=18$) to the silicates, chlorides, and sulfates found in paleo-detectors. We typically have $Z_1\ll Z_2$, which reduces the impact of screening and enhances the stopping power \citep{Sigmund:1997}. However, these results generally hold for velocities much greater than the Bohr velocity, $e^2/(4\pi\varepsilon_0\hbar)\simeq7\times10^{-3}c$. Our fiducial velocity is $\sim10^{-3}c$, and the effective charge may be different in this limit. \citet{Brandt:1982} find that the effective charge increases at low velocities, so we expect that the aforementioned effective charge of $Z_{\mathrm{eff}}\simeq 0.7Z_1$ is an estimate of a lower bound rather than an upper bound.

Since the stopping power of bare CPRs is very large for detection purposes, screening effects of this order have no qualitative impact on detection signatures. Thus, in subsequent calculations, we will ignore the distinction between bare and atomic CPRs. In particular, we will assume that the ionization and scintillation yields are comparable for the two projectiles. Note that we do not consider excitation or ionization of the CPR atom, as these effects would tend to increase the stopping power still further.

\section{Constraints and future prospects}
\label{sec:prospects}

\subsection{MACRO experiment}
\label{sec:macro}
The MACRO experiment \citep{Ambrosio:2002qq} performed a search for GUT-scale magnetic monopoles, placing an upper limit on the monopole flux at $\SI{5.5e-4}{\meter^{-2}.\year^{-1}}$ (90\% CL). This is significantly lower than the flux we estimate in \cref{eq:flux-estimate}. Since monopole transits are similar in many respects to CPR transits, we investigate the extent to which this bound can be applied to the CPR population.

The MACRO experiment consists of six independent analyses with different experimental properties. At velocities $\beta\sim10^{-3}$, there are two applicable constraints: the Wave Form Digitizer (WFD) analysis of the liquid scintillator system, and an analysis of streamer tubes filled with a mixture of helium and n-pentane. The streamer tube analysis relies on the Drell effect \citep{Drell:1982zy}: an incident magnetic monopole excites helium atoms, which then ionize n-pentane molecules. The Drell effect is specific to magnetic monopoles, and we do not expect a comparable phenomenon to occur in the case of CPR transits unless they are also magnetically charged. Thus, we turn our attention to the WFD search.

The WFD analysis was based on a data from a liquid scintillator detector equipped with photomultiplier tubes. The trigger was designed for slowly-moving monopoles, and was sensitive to photoelectrons emitted in sequence over several microseconds. Transient signals with a duration below $\SI{100}{\nano\second}$ were discarded. The WFD search set an upper limit to the flux at $\SI{9.9e-4}{\meter^{-2}.\year^{-1}}$ (90\% CL). If we assume that the search was fully sensitive to CPRs, this corresponds to a bound on the charged fraction of $f\lesssim0.5\%$, sufficient to rule out CPRs as the dominant component of dark matter. A dedicated search in comparable hardware would certainly be capable of establishing a bound at least this strong.

\subsection{ICARUS experiment}
\label{sec:icarus}
We now restrict our attention to liquid argon detectors, of the type pioneered by \citet{Rubbia:1977zz}. For concreteness, we consider the ICARUS detector, which operated for three years at Gran Sasso Laboratory \citep{Antonello:2015zea} and is currently being installed as a short baseline neutrino detector at Fermilab \citep{Acciarri:2016ooe}.

ICARUS consists of two chambers, each containing approximately 480 tons of liquid argon. The chambers are equipped with photomultiplier tubes and wire cages, so they are capable of detecting both scintillation light \citep{Gastler:2010sc} and ionization \citep{Bondar:2014nra} from keV-scale nuclear recoils. For the majority of the data collection at Gran Sasso, ICARUS was only triggered in time coincidence with the CERN neutrino beam, and therefore would likely have ignored any potential signal from a CPR transit.

The mean scintillation efficiency for nuclear recoils in liquid argon is about 0.25 \citep{Gastler:2010sc}, corresponding to a scintillation yield of approximately 13 photons per keV of deposited energy. We simulated the passage of a CPR through liquid argon with \SRIM~and found a stopping power per unit target density of approximately \SI{93}{\mega\electronvolt/(\gram/\centi\meter^2)}. Even if we only include recoils above \SI{10}{\kilo\electronvolt}, the resulting scintillation yield is $\sim\SI{7e5}{photons/\centi\meter}$. Using a conservative estimate of \SI{6}{e^-/\kilo\electronvolt} for the secondary ionization yield \citep{Bondar:2014nra}, we find that the transit yields $\sim\SI{3e5}{e^-/\centi\meter}$. Both signals are well above the detection thresholds of ICARUS' readout electronics. Indeed, these signals are much more significant than the scintillation and ionization signals from minimum-ionizing particles like muons, which have a stopping power of about \SI{1.5}{\mega\electronvolt/(\gram/\centi\meter^2)}.

\begin{figure}
    \centering
    \includegraphics[width=0.7\columnwidth]{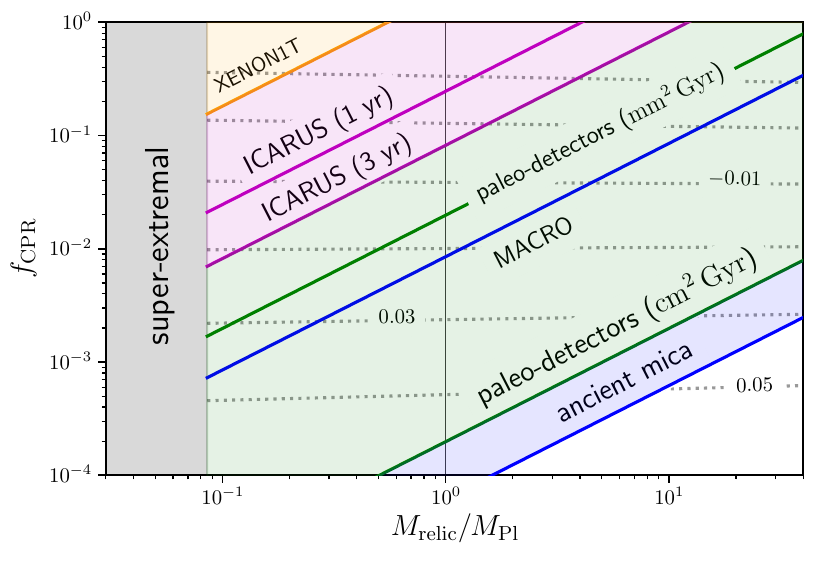}
    \caption{
        Projected 99\% CL upper limit on the mass and density of CPRs with experiments of several classes. See text for details. \textbf{Orange:} a \SI{10}{\year} exposure of XENON1T \citep{Aprile:2018dbl}. \textbf{Magenta:} solid: a \SI{3}{\year} exposure of ICARUS. The dashed line shows a \SI{1}{\year} exposure. \textbf{Green:} solid: estimated limits from a paleo-detector with $\mathcal E=1$ and a \SI{1}{\centi\meter^2.\giga\year} exposure. The dashed line shows a \SI{1}{\milli\meter^2.\giga\year} exposure. \textbf{Blue:} strongest possible limits from monopole searches, including a direct search by MACRO and a search for tracks in ancient mica \citep{Ghosh:1990}. (\cref{sec:macro}). \textbf{Dotted gray:} relic fractions produced assuming an initial a power-law mass function with index $\gamma$. Contours step from $\gamma=-0.05$ to $\gamma=0.05$ from top to bottom in increments of $0.02$. \textbf{Shaded gray:} region prohibited by super-extremality for a charge of $1e$.
    }
    \label{fig:constraints}
\end{figure}

The combination of high scintillation yields, high ionization yields, and a long crossing time (several \SI{}{\micro\second}) in the detector would be a smoking-gun signature of a CPR transit. A dedicated search should be able to use these factors to discriminate against cosmic ray backgrounds, even with ICARUS at the surface. In \cref{fig:constraints}, we show the expected upper limits on the CPR density that could be obtained from a dedicated search with ICARUS on various time scales. The limits are presented as a function of the fraction of dark matter in the form of detectable charged relics.

While we evaluate fiducial constraints using the parameters of the ICARUS experiment, there are several other similar liquid argon detectors that may be also be usable for CPR searches. Assuming that the readout electronics of each detector are sensitive to CPR transits, the maximum abundance of CPRs compatible with non-detection scales inversely with the effective area $A_{\mathrm{eff}}$ of the detector, defined as the cross-sectional area of the detector averaged over the arrival direction. For a rectangular prism with side lengths $L_i$, we have $A_{\mathrm{eff}}=\frac12(L_xL_y+L_xL_z+L_yL_z)$, and for a cylinder of radius $r$ and length $L$, we have $A_{\mathrm{eff}}=\frac12\pi r(L+r)$. We summarize the effective areas of various detectors in \cref{tab:effective-areas}.

\begin{table}\centering
    {
        \small
        \begin{tabular}{|c|c|c|c|c|c|c|}
            \hline
                ICARUS &
                ArgoNeuT &
                LArIAT &
                SBND &
                ProtoDUNE &
                DUNE &
                MicroBooNE
            \\\hline
                80.5 \citep{Rubbia:2011ft} &
                1.01 \citep{Anderson:2012vc} &
                1.01 \citep{Cavanna:2014iqa} &
                22.4 \citep{Antonello:2015lea} &
                104 \citep{Abi:2017aow} &
                792 \citep{Acciarri:2016ooe} &
                28.3 \citep{Acciarri:2016smi}
            \\\hline
        \end{tabular}
    }
    \caption{\normalsize Effective areas of current and future liquid argon detectors in \SI{}{\meter^2} (see text for details). The strength of constraints scales linearly with the detector area. Of these, ICARUS is the largest detector currently operational.}
    \label{tab:effective-areas}
\end{table}

\subsection{Paleo-detectors}
\label{sec:paleo-detectors}
We now examine the prospects for detecting CPRs with paleo-detectors \citep{Baum:2018tfw,Drukier:2018pdy,Edwards:2019puy}. For WIMP detection, the key prediction is the spectrum of lengths of the ionization tracks produced by recoiling nuclei. For our purposes, the CPR itself takes the role of the nucleus. Since a CPR is negligibly slowed even by macroscopic volumes of matter, an ionization track from a CPR transit will be extremely long. Recall that we are interested in typical kinetic energies of order \SI{e21}{\electronvolt}. If we take the typical stopping power in rock to be $\du E/\du x\sim\SI{100}{\mega\electronvolt/\centi\meter}$, as in \cref{eq:linear-stopping-estimate}, then CPRs should pass through Earth entirely without losing more than a fraction $\sim10^{-4}$ of their energy. In principle, the resulting tracks would cross the entire planet, although they would be disrupted over time by geological effects.

Furthermore, the exposure time is such that a paleo-detector with cross-sectional area $A$ should have a number of ionization tracks given by
\begin{equation}
    N \sim 200\left(
        \frac{A}{\SI{1}{\milli\meter^2}}
    \right)\left(
        \frac{t_{\mathrm{obs}}}{\SI{e9}{\year}}
    \right)\mathcal E.
\end{equation}
where the efficiency $\mathcal E$ accounts for the probability of track production, track survival, and track detection. If $\mathcal E\sim1$, this extremely high track density means that paleo-detectors should be capable of placing very stringent constraints on the CPR population. For this reason, track searches in ancient minerals have already been used to constrain the flux of supermassive magnetic minerals, which have a similar detection signature \citep{Ghosh:1990}. If these searches are taken to be sufficiently sensitive to detect a CPR track, then we can already infer a limit of $\sim\SI{4e-6}{\meter^{-2}.\year^{-1}}$ on the flux of CPRs.

The disadvantage of paleo-detectors is that a particular track cannot be identified as a CPR in a small volume of material. Paleo-detectors do not directly measure the speed of the transit, which is an important experimental signature for our purposes. In a large piece of material undisturbed by geological processes, a CPR transit may be identifiable by the length of the track, but this may require additional technological development and other modifications to paleo-detector-based searches. Thus, paleo-detectors can constrain CPRs, but may not easily furnish a confirmed detection. Still, we show prospective constraints from paleo-detectors in \cref{fig:constraints}.

\section{Discussion}
\label{sec:discussion}
In the foregoing sections, we have argued that Planck-scale relics of evaporating primordial black holes may generically have charges of order $e$, and we have shown that plausible forms of the PBH mass spectrum lead to a significant CPR population today. The formation of these objects is inextricably connected to quantum gravity, and the process is sensitive to new physics at extremely high energies. We have further shown that if CPRs constitute a significant fraction of dark matter, then they can be detected terrestrially. Indeed, not only are such objects detectable, but the detection signature would be a smoking gun with few alternative possibilities.

The implications of such a detection cannot be overstated. In addition to furnishing a direct detection of dark matter, this would confirm the PBH paradigm, providing great insight into the conditions of the early universe. An abundant population of such objects would furnish the first system for the direct laboratory study of gravity in the quantum regime. Of course, most immediately, even a single detection would establish that black holes do not evaporate completely, but leave behind a relic. Even non-detection may provide significant information: if dark matter is one day found to be composed of PBHs, and their mass function is established, the fraction of CPRs produced is easily calculated. Non-detection at that level would establish that evaporating black holes leave no relics, or that such relics cannot be charged.

With these objectives in mind, we have tentatively derived existing constraints from the MACRO experiment and ancient mineral searches, which already exclude $\fcpr=1$ at the 99\% confidence level across the entire mass range we consider. However, there are numerous scenarios which predict a much smaller CPR fraction. If dark matter is composed mainly of PBHs at a higher mass scale, but produced with a broad mass spectrum, then a low-mass tail evaporates to the Planck scale, producing a smaller abundance of CPRs. Further, the uncertainties in the estimation of the charge fraction imply that only a small fraction of Planck-scale relics may be charged. Thus, there is ample motivation to search for smaller CPR fractions.

Fortunately, performing such a search requires no new equipment apart from possible modifications to experimental triggers. We have projected stringent constraints from the ICARUS experiment, which can set a bound $\fcpr\lesssim 10^{-2}$ if $\mstop\simeq\mpl$. Our projected constraints from paleo-detectors strengthen the bound to $\fcpr\lesssim10^{-4}$ at $\mpl$, and can constrain the CPR dark matter fraction at the per cent level even if $\mstop$ lies an order of magnitude above $\mpl$. Taken together, these bounds would be sufficient to exclude CPRs as a significant fraction of dark matter even if they lie at a different mass, or acquire spontaneous charge with a somewhat smaller probability. They are also extremely inexpensive to obtain.

\section{Conclusions}
\label{sec:conclusions}
Primordial black hole dark matter remains a viable and parsimonious dark matter candidate. If dark matter is in the form of Planck-scale relics, such objects would be effectively sterile with respect to the standard model if neutral. In this work, we have argued that such relics may in fact carry charges of order $e$, in which case dark matter could be composed largely of Charged Planck-scale Relics (CPRs). We have shown that CPR dark matter is detectable terrestrially, with initial constraints already set by the null results of monopole searches. Moreover, upcoming experiments can be used to conduct a much more sensitive search for CPRs with little or no modification. Even a single detection would come with significant implications for black hole physics, the behavior of gravity in the quantum regime, and the nature of dark matter.

The interpretation of non-detection is more subtle. Optimistically, null results can constrain the overall population of relic black holes, with implications for either the PBH mass function or the quantum gravity mechanisms that stabilize them. Realistically, however, the argument we present in this work only motivates the possibility of a substantial charged fraction---it is not a rigorous prediction. As such, we cannot immediately draw conclusions regarding the abundance of black hole relics in general.

However, up to some of the other uncertainties discussed in this work, it is conceivable that the charge distribution could be rigorously predicted within the context of a candidate quantum gravity theory. In this case, the abundance and charge distribution of relic black holes would become testable predictions of such a theory. Remarkably, as we have shown, we may already have experimental access to such a scenario. The constraints we draw on the abundance of CPRs may thus translate into constraints on the structure of physics at the Planck scale.

\acknowledgments
BVL and SP are partially supported by the U.S. Department of Energy grant number DE-SC0010107. We thank Bruce Schumm, Jahred Adelman, Anthony Aguirre, Louise Suter, Bill Atwood, Robert Johnson, Christian Byrnes and Philippa Cole for valuable conversations. We are especially grateful for conversations with Steven Ritz, including the suggestion to evaluate liquid argon detectors.

\bibliographystyle{unsrtnat}
\bibliography{main}

\end{document}